# Seismic potential map of Greece calculated for the years 2005 and 2010. Its correlation to the large (Ms>=6.0R) seismic events of the 2000 – 2009 period.

Thanassoulas[1], C., Klentos[2], V.

1. Retired from the Institute for Geology and Mineral Exploration (IGME), Geophysical Department, Athens, Greece.
   e-mail: thandin@otenet.gr - URL: www.earthquakeprediction.gr

2. Athens Water Supply & Sewerage Company (EYDAP),
   e-mail: klenvas@mycosmos.gr - URL: www.earthquakeprediction.gr

**Abstract.**

The seismic potential maps for Greece, particularly for the years 2005 and 2010 (end of 2009), have been calculated following the methodology of the "lithospheric seismic energy flow model". The compiled, for the year 2005, map is compared to the observed large seismicity of 2005 – 2010 period. Furthermore, a comparison is made of the seismic charge status regarding the year 2005 to the one of 2000. It is revealed that there exists an increase of the seismic potential which results into triggering, in the next five years period (2005 – 2010), almost double the number of large EQs, compared to the ones observed during 2000 - 2005 period of time. Finally, estimation is made about the seismic accelerating deformation status of the entire Greek territory. It is shown that, since 2004, accelerated deformation monotonically increases and it is speculated that some large seismic events is possible to occur within the next 1 - 3 years. The average "virtual" magnitude of these seismic events, considering the entire Greek territory as a single unit seismogenic area, has been calculated as Ms = 8.18R. In practice, these events will be decomposed into a number of smaller in magnitude, but still large too, seismic events located at normally different seismogenic areas.

## 1. Introduction.

Generally, the term "seismic hazard" of a region refers to a quantity "**H**" which depends on the expected seismic motion intensity of that region. The seismic hazard is usually expressed in terms of ground acceleration, ground velocity, ground dislocation or as macroseismic intensity (Papazachos et al. 1989).

Such a typical seismic hazard map for the Greek territory was compiled by Papazachos et al. (1989) which is presented in the following figure (1) at left. The entire Greek territory has been divided into four different zones, scaling from (I) to (IV), with the (I) scale being the one with the lowest value of seismic hazard.

Later on, after the occurrence of some large EQs in Greece, the seismic hazard map of 1989 was modified by OASP (2004) and the new seismic hazard map consists of three zones only. That map is presented in the middle of figure (1).

The seismic hazard of an area mainly depends upon its seismicity. Therefore, the variability of the seismicity of an area, at some time scale, suggests variable seismic hazard. The latter indicates the necessity of a more frequent seismic hazard map compilation or the use of the so called "seismic potential map" (Thanassoulas et al. 2003, Thanassoulas 2007). What is presented in the seismic potential map is the spatial distribution of the seismic energy (in terms of Ms) stored in the analyzed area. Consequently, what is presented is the maximum expected magnitude of an earthquake, at any place of the map, when it occurs. In figure (1) at right, is presented the seismic potential map of Greece calculated for the year 2000.

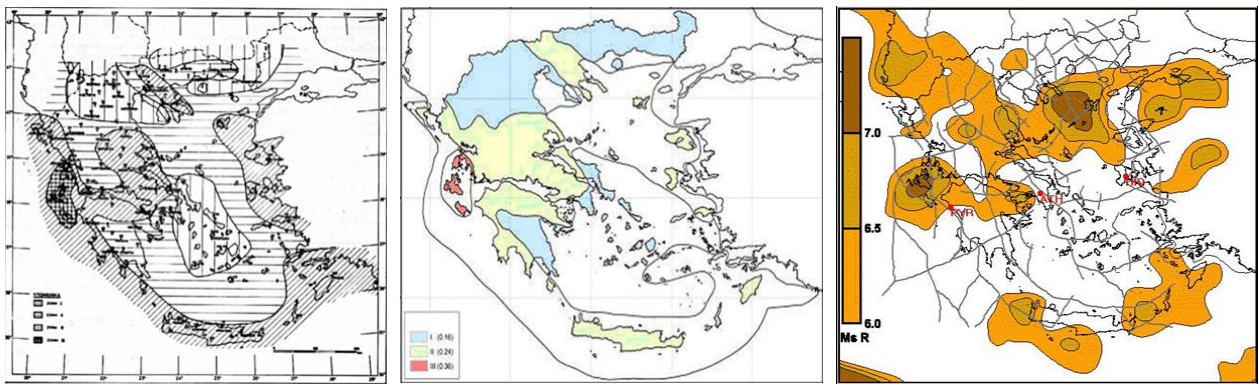

Fig. 1. Left: Map of ground acceleration ( Papazachos 1989). Middle: its recent revision (OASP, 2004) and Right: seismic potential map for the year 2000 (Thanassoulas et al. 2003, Thanassoulas 2007).

The seismic potential map was compiled with a calculated earthquake lower magnitude threshold of Ms = 6.0 R and was compared to the corresponding followed up seismicity. It was found that decreased in size extent seismically charged areas (brown colored areas) were followed by decreased large (Ms>6.0R) seismicity. Moreover, an increase of the stored seismic energy corresponding to only a half (0.5R) of a unit of the Richter scale, which resulted in to a wider colored area, corresponded to a large increase of the large earthquake seismicity, which coincided quite well to the suggested by the seismic potential map seismic energy charged areas (Thanassoulas 2008, 2008a, 2008b).

In this work we calculated the seismic potential maps of Greece for the years 2005 and 2010 (end of 2009) and compare them to the seismicity that took place in the period of time from the year 2000 until start of the year 2010.

## 2. Map compilation and analysis.

The earthquake data, which we used for this work, were downloaded from the web site of the National Observatory of Athens. The earthquake catalog which covers the period 1901 to 2009 was used for the compilation of both seismic potential maps. The processing of the data was performed by using the methodology which was presented by Thanassoulas et al. (2001, 2003) and Thanassoulas (2007, 2008b).

In the following figure (2) the compiled maps are presented.

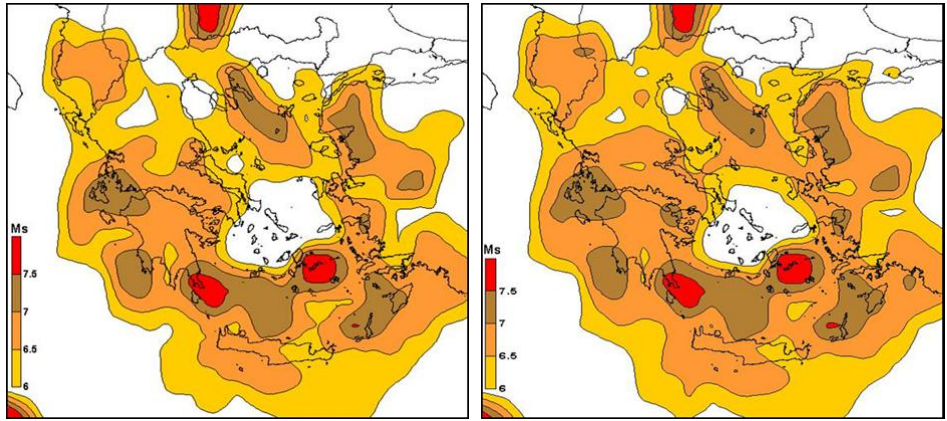

**Fig. 2.** Calculated seismic potential maps of Greece for the years 2005 (left) and 2010 (right). Colored areas indicate the maximum EQ magnitude (in Ms) expected at each case (see corresponding color bar).

At a first glance, there is no significant difference between these two maps. However, a closer inspection indicates that the 2010 map suggests a rather larger degree of seismic energy charge. This is more evident if only the range of 7.0R>Ms>6.5R is being considered. In order to evaluate the compiled maps, these will be compared to the seismicity of the following 5 years period with various seismic magnitude minimum thresholds and for different seismic potential lower thresholds. At the background of each map, simultaneously, are presented, as gray thick lines, the deep lithospheric fracture zones deduced by the analysis of the corresponding gravity field (Thanassoulas 1998, 2007).

For each map a (**P**) value will be calculated which indicates the percentage (**P**in / **P**tot) of the total number **P**in of the EQs that did occur in the indicated seismically charged zones over the **P**tot of the total number of large EQs that took place all over the Greek territory.

That is $P = P_{in} / P_{tot}$. (1)

Following in figure (3, 4, 5) is analyzed the seismic potential map of year 2005.

**Seismicity lower threshold = 6.0R.**

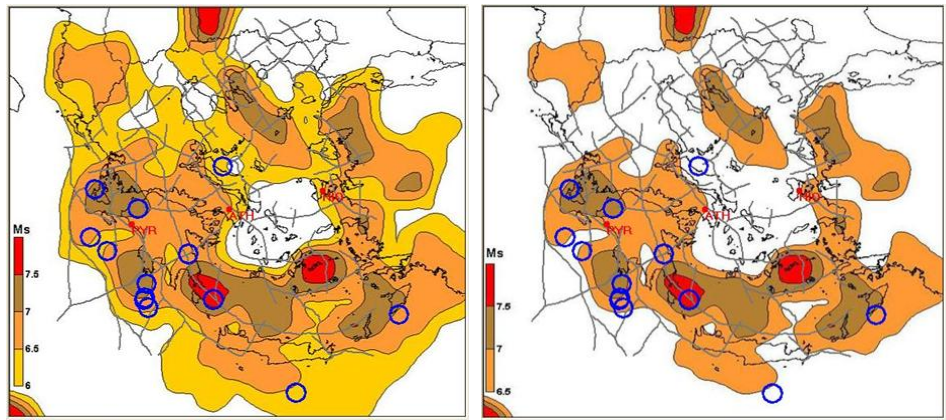

**P = 13 / 13 = 1.0 or 100%**          **P = 12 / 13 = .923 or 92.3%**

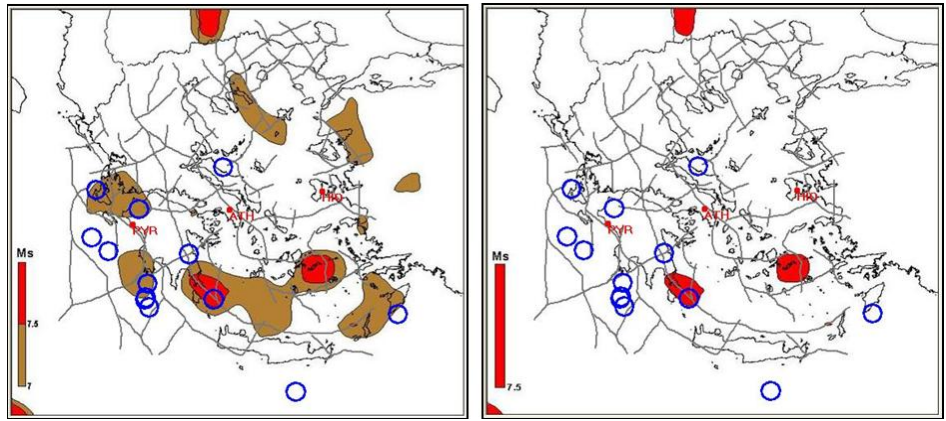

**P = 7 / 13 = 0.538 or 53.8%**   **P = 1 / 13 = .077 or 7.7%**

**Fig. 3. Seismic potential maps of the year 2005 compared with the corresponding seismicity (Ms >= 6.0R) for the period of 2006 – 2009. Seismic potential lower threshold: upper left = 6.0R, upper right = 6.5R, lower left = 7.0R, lower right = 7.5R.**

**Seismicity lower threshold = 6.5R.**

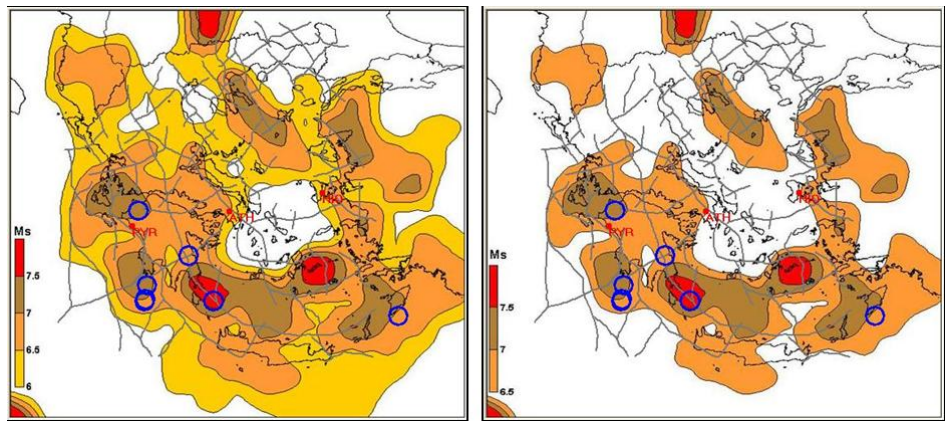

**P = 7 / 7 = 1.00 or 100%**   **P = 7 / 7 = 1.00 or 100%**

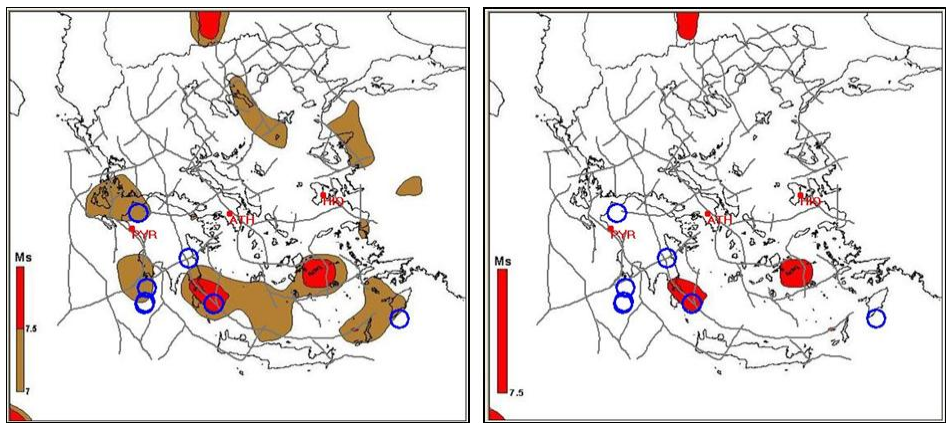

**P = 7 / 7 = 1.00 or 100%**   **P = 1 / 7 = 0.1428 or 14.28%**

**Fig. 4. Seismic potential maps of the year 2005 compared with the corresponding seismicity (Ms >= 6.5R) for the period of 2006 – 2009. Seismic potential lower threshold: upper left = 6.0R, upper right = 6.5R, lower left = 7.0R, lower right = 7.5R.**

**Seismicity lower threshold = 7.0R.**

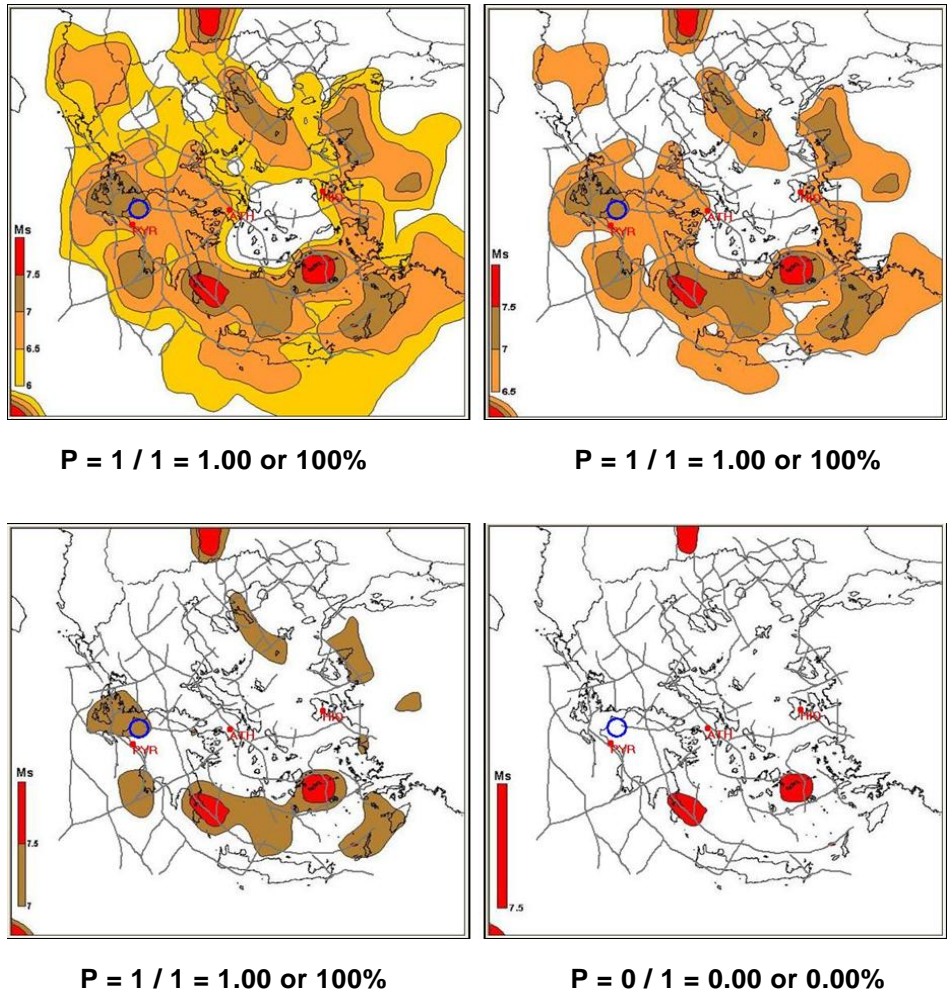

P = 1 / 1 = 1.00 or 100%        P = 1 / 1 = 1.00 or 100%

P = 1 / 1 = 1.00 or 100%        P = 0 / 1 = 0.00 or 0.00%

**Fig. 5. Seismic potential maps of the year 2005 compared with the corresponding seismicity (Ms >= 7.0R) for the period of 2006 – 2009. Seismic potential lower threshold: upper left = 6.0R, upper right = 6.5R, lower left = 7.0R, lower right = 7.5R.**

As it is evident, there are no earthquake data to be compared with the seismic potential map of the year 2010. Obviously such a comparison can be utilized by any interested researcher in the future after the year 2015. In the mean time we can compare the specific map of 2010 to the large seismicity which took place during the past ten (10) years. This is demonstrated in the following figures (6, 7, 8).

**Seismicity lower threshold = 6.0R.**

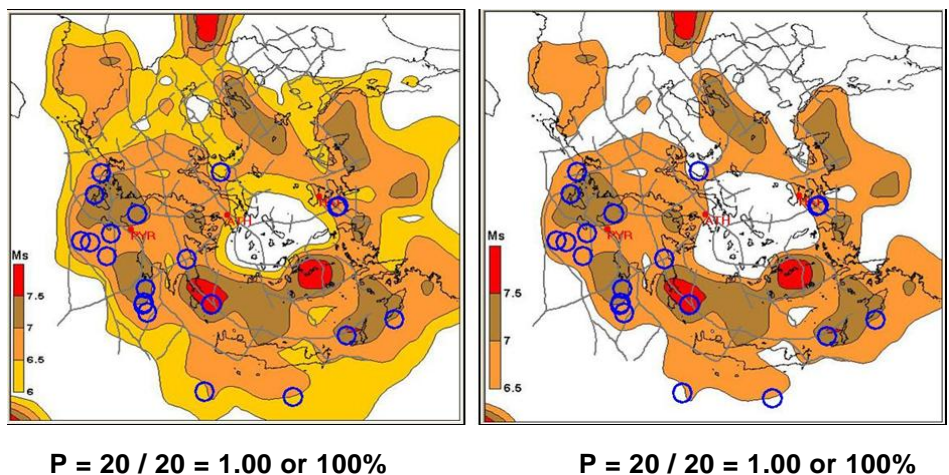

P = 20 / 20 = 1.00 or 100%        P = 20 / 20 = 1.00 or 100%

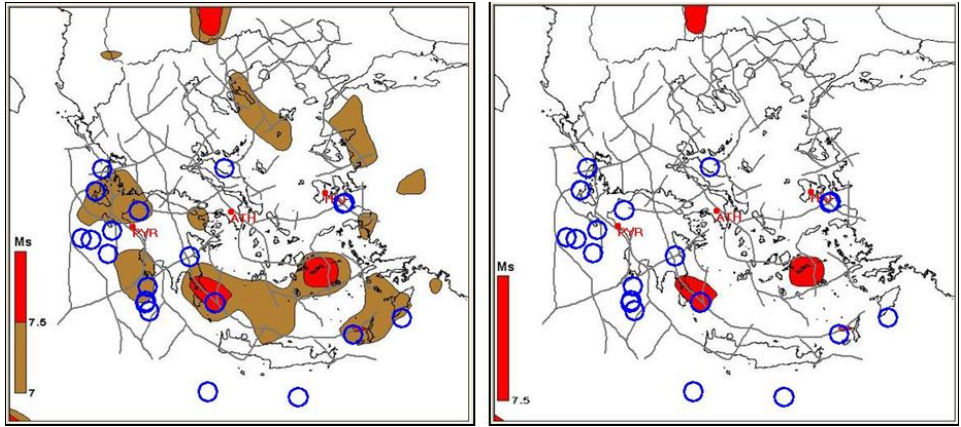

| P = 13 / 20 = 0.65 or 65% | P = 1 / 20 = 0.05 or 5% |

**Fig. 6. Seismic potential maps of the year 2010 compared with the corresponding seismicity (Ms >= 6.0R) for the period of 2000 – 2009. Seismic potential lower threshold: upper left = 6.0R, upper right = 6.5R, lower left = 7.0R, lower right = 7.5R.**

**Seismicity lower threshold = 6.5R.**

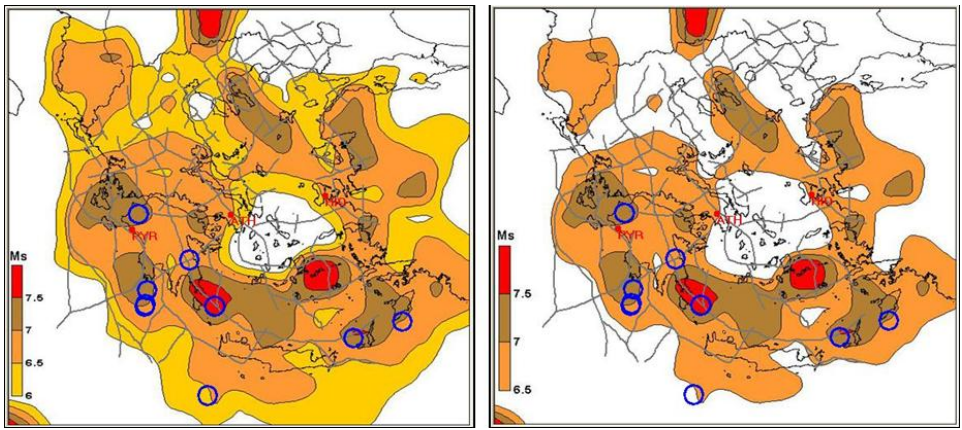

| P = 9 / 9 = 1.00 or 100% | P = 9 / 9 = 1.00 or 100% |

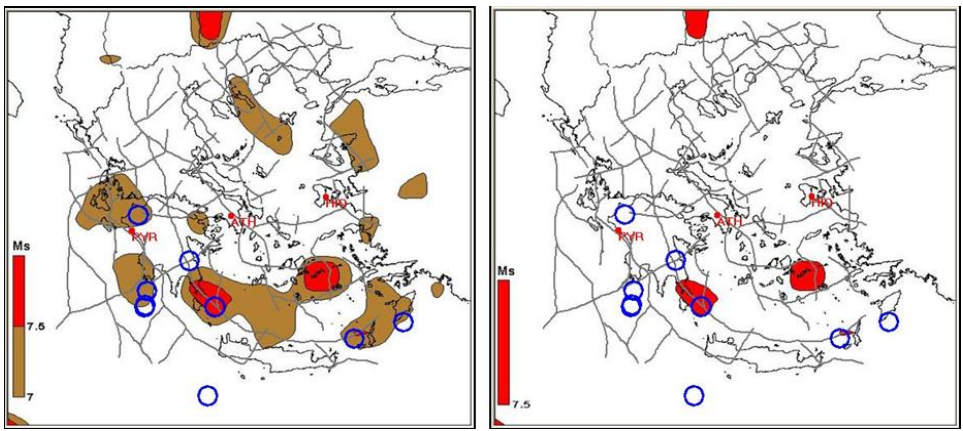

| P = 7 / 9 = 0.78 or 78% | P = 1 / 9 = 0.11 or 1.10% |

**Fig. 7. Seismic potential maps of the year 2010 compared with the corresponding seismicity (Ms >= 6.5R) for the period of 2000 – 2009. Seismic potential lower threshold: upper left = 6.0R, upper right = 6.5R, lower left = 7.0R, lower right = 7.5R.**

**Seismicity lower threshold = 7.0R.**

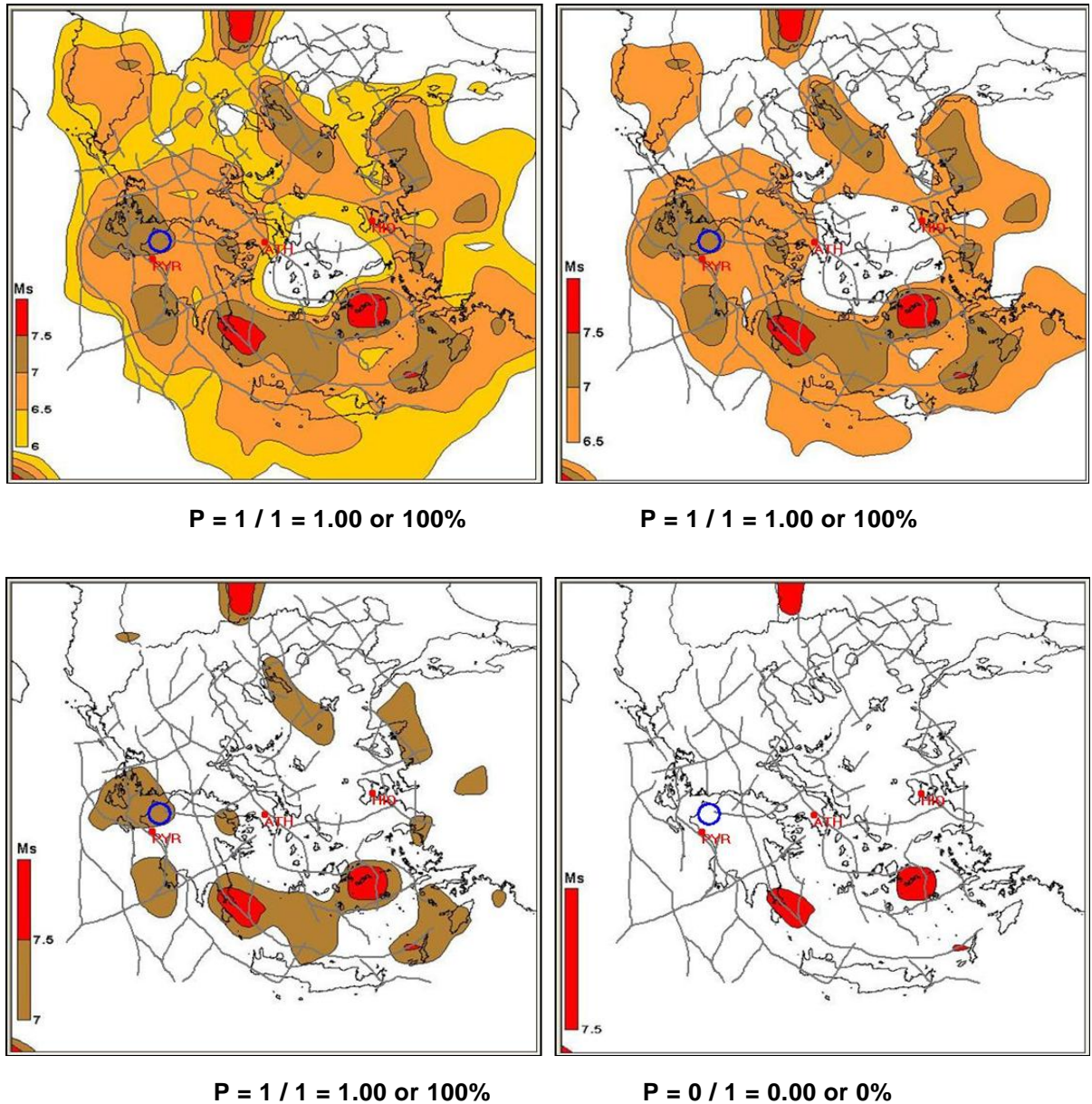

Fig. 8. Seismic potential maps of the year 2010 compared with the corresponding seismicity (Ms >= 7.0R) for the period of 2000 – 2009. Seismic potential lower threshold: upper left = 6.0R, upper right = 6.5R, lower left = 7.0R, lower right = 7.5R.

## 3. Discussion – Conclusions

In the recent studies of Thanasssoulas et al. (2003) and Thanassoulas (2007, 2008b), regarding the seismic potential of the Greek territory, it was shown that the latter changes quite fast within a five years period of time. Consequently, the seismic hazard map of Greece, calculated for any time, is valid only for a rather short time period. Therefore, it is suggested to use, instead of the traditional seismologically compiled seismic hazard map, the seismic potential map, that is a more direct way to evaluate the seismic hazard of a study area.

At this point it is worth to recall the large (Ms>6.0R) seismicity that followed, the next five years period of time (2001-2005), after the seismic potential map was calculated for the year 2000 and to compare it with the one of 2005 and its subsequent large seismicity of the next five years (2005 – 2009). This is demonstrated in the following figure (9).

**Map of year 2000 (EQs of 2000 – 2005)**  **Map of Year 2005 (EQs of 2005 – 2009)**

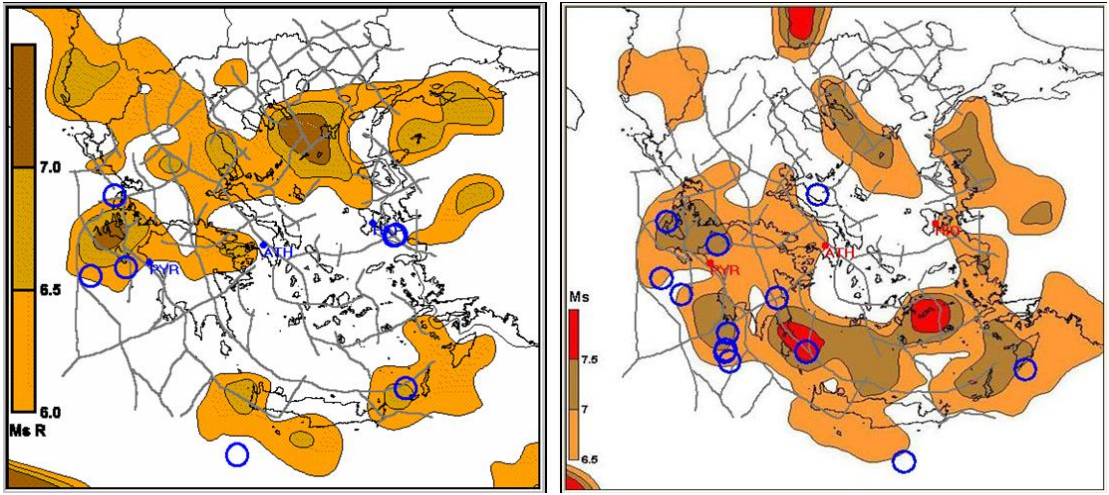

P = 4 / 7 = 0.57 or 57%        P = 12 / 13 = .923 or 92.3%
P = 6 / 7 = 0.86 or 86%

Fig. 9. Comparison of the seismic potential maps of 2000 (left) and 2005 (right) with the large (Ms>6.0R) seismicity that took place during the periods 2000 – 2005 and 2005 – 2009 (end). The second (lower line) P value of 2000 map is obtained after accepting the double seismic event at eastern Greece as of a marginal hit.

What is very interesting in figure (9) is the fact that an increase for 0.5R of the observed seismic potential from 2000 to 2005 resulted into triggering almost double the number of large EQs (from 7 to 13) within the next five years period while the hit rate (EQs in the predefined areas) remained at a high level (57.0% - 92.3% ).

A different example of such a change in the seismic potential is presented by comparing the seismic potential map of Greece compiled for the year 2000 to the one compiled for the year 2010. The latter is presented in the following figure (10).

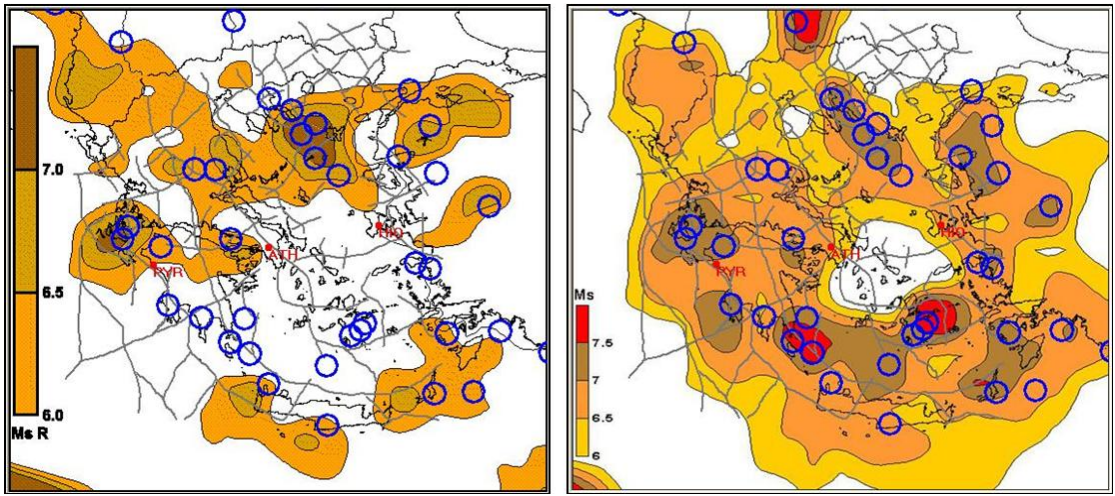

Fig. 10. Comparison of the seismic potential map of Greece compiled for the year 2000 (left) and the year 2010 (right). EQs of Ms>=6.8R for 1901 – 2009 are shown in blue circles.

It is made clear that within 10 years period of time a rather large increase of the seismic potential occurred. This had as an effect the occurrence of 20 large EQs (Ms > 6.0R) for the same time period. The seismic potential levels of 7.0R > Ms > 6.5R exhibit the largest seismicity effect. The corresponding hit rate (EQs within the specified seismic potential area) of the occurred large EQs ranges from 53.8% to 100% which we consider as quite satisfactory.

A more general comparison is made of the seismic potential maps compiled for the years 2000 and 2010 (fig. 10) to the large (Ms>6.8R) seismicity of the period 1901 – 2009. It is clearly shown that the majority of the large seismicity takes place in the dark-brown and red colored seismic potential areas (Ms>7.0R) while at the same time highly seismic potential charged areas are prescribed quite well by clusters of large EQs (fig. 10 right).

The observed seismic potential charge difference, between the map of year 2000 to the map of year 2010, suggests that the Greek territory, during these last 10 years, is being charged continuously which in turn indicates that the observed in the recent years large seismicity has not come yet to an end.

A more remarkable feature of the seismic potential maps of 2005 and 2010 is their cyclic character. It is clearly shown that the seismic potential decreases drastically in the centre of the Aegean plate (white colored area). A comparison of a lower level seismicity of Ms = 4.5R, for the entire period of 1901 to 2009, to the seismic potential map of 2010 results into a very good agreement. The latter is shown in the following figure (11).

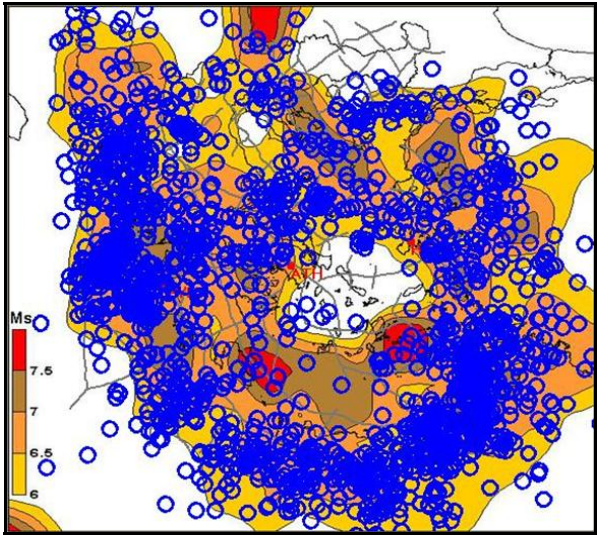

**Fig. 11.** Comparison of the seismic potential map of Greece compiled for the year 2010 to the corresponding seismicity (blue circles) of Ms = 4.5R for the time period of 1901 to 2009.

The case of the Kythira EQ (Ms = 6.4R, January 2006) is a very good example of the use of the seismic potential maps. The seismic potential map of year 2005, with a minimum threshold of 7.5R shown in figure (12), indicated three "red" narrow zones in the entire Greek territory. The first one is located at north, at the Greek – Bulgarian frontiers and the other two are located at the southern part of Greece. The one located at the left is the Kythira area. The blue circle indicates the seismologically calculated epicenter of the same EQ. Therefore, if this map had been calculated at the right time (end of 2005) it could be possible to have been warned about the seismic status of this seismogenic region in Greece.

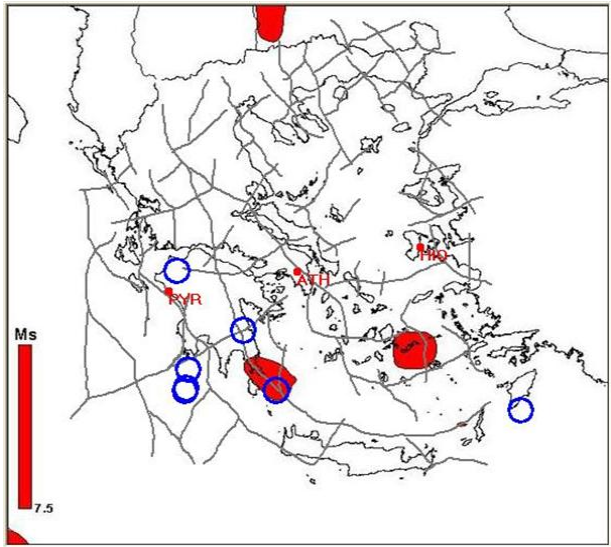

**Fig. 12.** Narrow areas of Greek territory (red color) charged (at 2005) with seismic energy capable to generate an EQ of Ms>7.5R. The Kythira EQ coincides to the southern left one of them.

It is important to point out that the Kythira seismogenic area is still highly charged (Thanassoulas, 2007) and capable to generate large EQs in the future. The very same charged condition is suggested by the seismic potential map of figure (12).

By comparing the seismic potential maps generated for the period of 1975 – 2000 to the similar maps compiled for the years 2005 and 2010 it is identified an observable increase of the seismic potential of the entire Greek territory. Therefore, we will consider the entire Greek territory as a single seismogenic unit region and will analyze it in terms of accelerated deformation. The method to be used is the one of the lithospheric seismic energy flow model postulated by Thanassoulas (2007, 2008a).

The cumulative seismic energy release in the entire Greek territory is presented as a function of time (one month sampling interval) in figure (13)

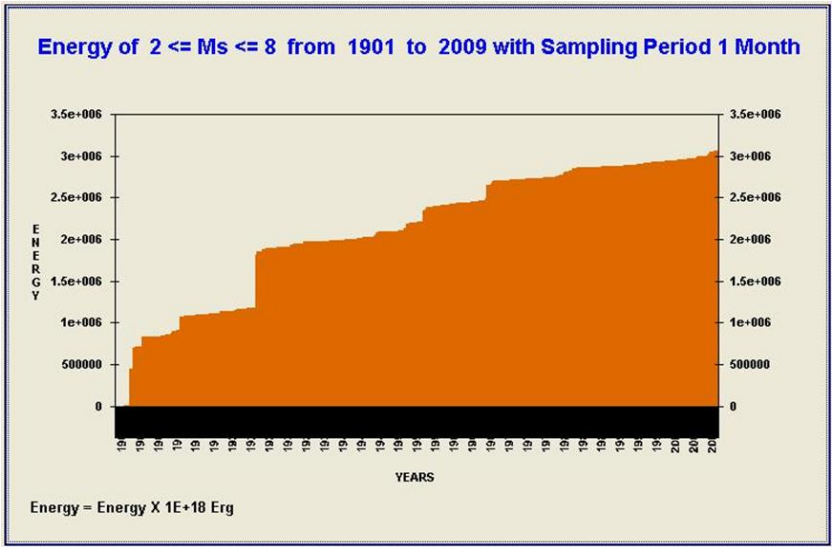

Fig. 13. Calculated cumulative seismic energy release in the entire Greek territory for the time period of 1901 to 2009.

From the previous graph it is possible to calculate the maximum expected magnitude of a future large EQ by choosing the appropriate "normal seismic energy release" function and its time of occurrence. The term "normal seismic energy release" means the one according to which no "lock" state exists in the seismogenic region and all the seismic energy release, theoretically, takes place through, a continuous in time, large number of small magnitude seismicity. Following are presented four different scenarios for the calculation of the maximum expected magnitude.

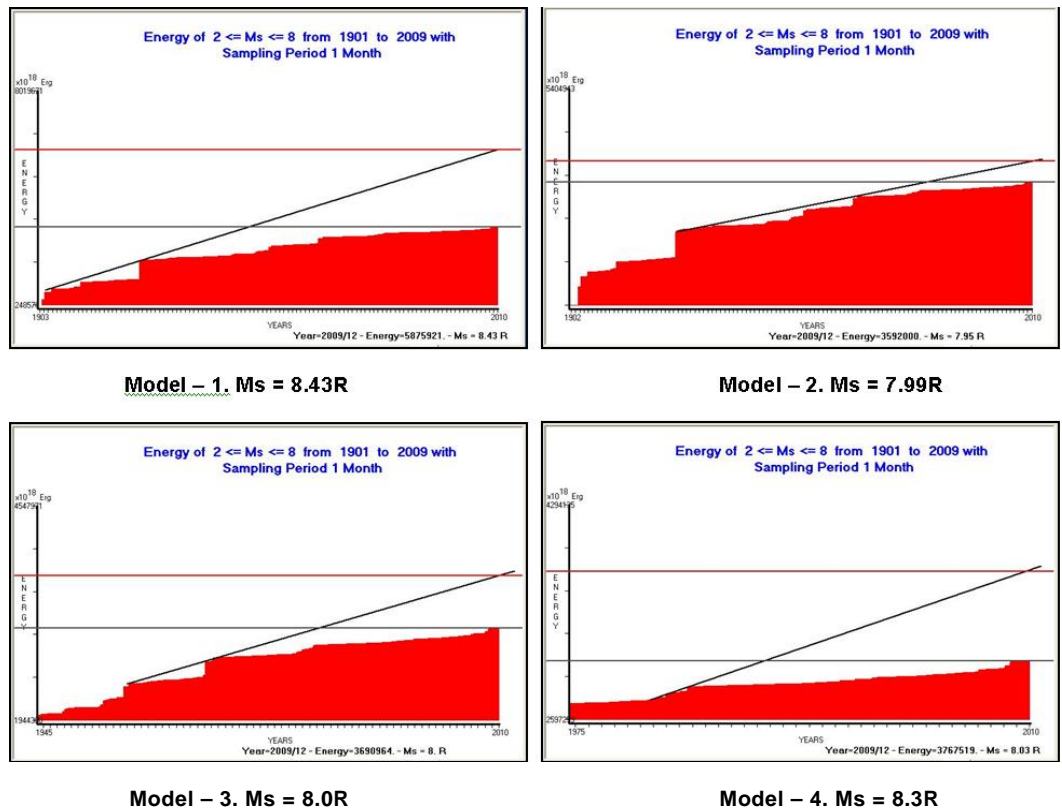

Model – 1. Ms = 8.43R        Model – 2. Ms = 7.99R

Model – 3. Ms = 8.0R         Model – 4. Ms = 8.3R

Fig. 14. Four different scenarios (models-1, 2, 3, 4) are presented for the calculation of the maximum expected magnitude (in Ms) of a large EQ in the Greek territory.

Although different features of the cumulative seismic energy release graph have been used, at different periods of time, the result is more or less the same. The expected maximum magnitude ranges from Ms = 7.99R to Ms = 8.43R with an average value of **Ms = 8.18R**. In practice, these "virtual" events will be decomposed into a number of smaller, but still large, seismic events which will take place, generally, at different seismogenic areas.

Further more we observe, from figure (13), that the Greek territory, since 1975, has entered a rather long "quiescence" period which has turn into an accelerated one after 1986. This is shown in the following figure (15) being a magnified part of figure (13).

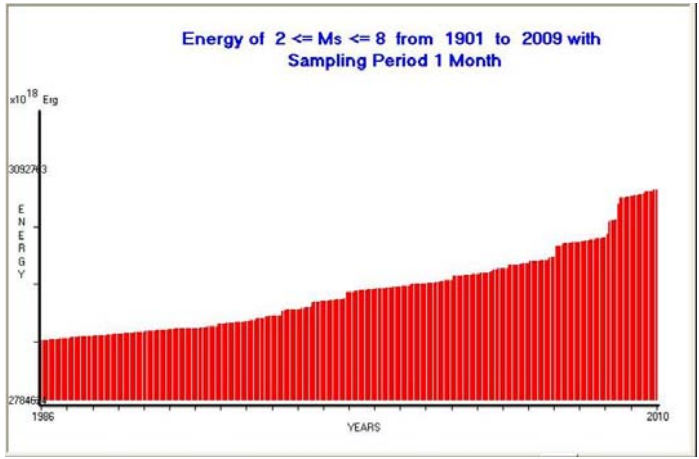

**Fig. 15.** Accelerated deformation observed for the period of 1986 to 2009.

A sixth order polynomial (fig. 16) has been fitted to the data of figure (15) for the following reasons. The first one is to make clearer the increase of the cumulative seismic energy release and the second is to utilize the analytical determination of the gradient (fig. 17) in time of the cumulative seismic energy release so that the accelerated deformation becomes more evident.

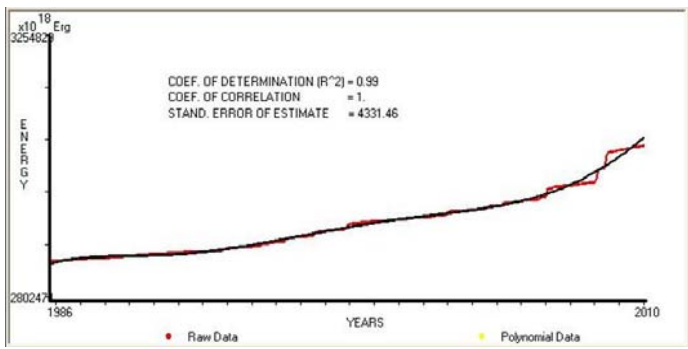

**Fig. 16.** $6^{th}$ order polynomial fitted on the data of the cumulative seismic energy release for the period from 1986 to 2010.

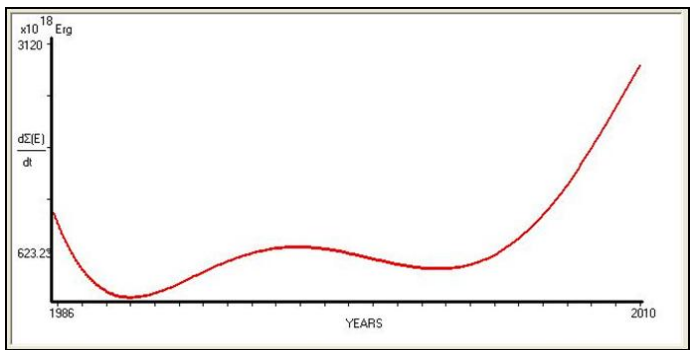

**Fig. 17.** Analytical time gradient of figure (16). It is clear that a considerable increase of the accelerated deformation has started since 2004 and is still increasing.

Similar cases, calculated for specific seismogenic areas in the past, which have been presented by Thanassoulas (2007), were followed by large EQs just after the calculated gradient achieved, as a graph, its sharper rise. For this case, and for the entire Greek territory, it is estimated that some rather large EQ is quite possible to take place within the next 1 – 3 years.

In conclusion, the seismic potential maps, which were calculated for the years 2005 and 2010, provided us with valuable information about the seismic status of the Greek territory, the specific seismogenic highly charged areas where large EQs do often occur and moreover inform us on what is possible to expect as maximum seismic activity in the entire Greek territory in the next years to come.